\documentclass[superscriptaddress,prl,twocolumn,showpacs,letterpaper,amssymb,amsmath,floatfix]{revtex4-1}
\pdfoutput=1

\usepackage{amsmath}
\usepackage{stmaryrd}
\usepackage{color}
\usepackage{graphicx}
\usepackage{mathrsfs}

\begin{document}

\title{Different roles  of Zn$^{2+}$ and Li$^{+}$ impurities in the CuO$_2$ plane in undoped cuprate compounds }

\author{Jia-Wei Mei}
\affiliation{Institute for Theoretical Physics, ETH Z\"urich, 8093 Z\"urich, Switzerland}

\date{\today}

\begin{abstract}
  A planar Mott insulator with easy plane N\'eel order can be mapped onto a Gutzwiller-projected spin Hall state. The in-plane substitution of Zn$^{2+}$ and Li$^{+}$ for Cu$^{2+}$  brings zero modes around the spin vacancy. The Gutzwiller projection precludes single particle excitations, however, preserves the zero modes which have a local spin moment for Zn$^{2+}$ and a local charged hole for Li$^{+}$, respectively. While the local spin moment for Zn$^{2+}$ is screened by the long-range spin correlations, the active charge degree of freedom for Li$^{+}$ impurity twists the spin background with long ranged disturbances. This proposal explains the very different roles of the Zn$^{2+}$ and Li$^{+}$ impurities on the magnetic perturbations in the CuO$_2$ plane in the undoped cuprate compounds.
\end{abstract}
\maketitle
The parent compound of high $T_c$ superconductors, e.g. La$_2$CuO$_{4}$, is antiferromagnetic (AF) insulator in which the long range AF order has the easy plane anisotropy in the CuO$_2$ plane due to the spin-orbit couplings\cite{Peters1988,Coffey1991,Yildirim1995}.  It is generally believed that the high $T_c$ superconductivity arises out of the coherent charge behavior of the dopant holes in the short range AF ordered spin background\cite{Anderson1987,Lee2006}. For La$_2$CuO$_4$,  a very small concentration of holes ($\sim 2\%$), introduced by Sr or Ba, suffices to suppress long range AF order and then the superconductivity appears immediately. The insulator-metal  transition is accompanied by  different competing magnetic and charge orders making it difficult to straighten out the magnetic disturbance from the dopant. The in-plane substitution of Zn$^{2+}$ and Li$^{+}$ for Cu$^{2+}$ provides the opportunity to study magnetic disturbances from the dopant without competing charge orders\cite{Alloul2009}. The Zn$^{2+}$ substitution brings the spin vacancy to the plane and disturbs the spin magnetism only with short ranged perturbations. The long range N\'eel order for the Zn$^{2+}$ substitution survives until the concentration close to the site dilution percolating threshold\cite{Vajk2002}.  Comparing with Zn$^{2+}$ substitution, Li$^+$ brings an extra hole with it. The Li$^+$ substitution shows a rapid suppression of long-range AF order like Sr but without a transition to a conducting state\cite{Rykov1995,Sarrao1996}. The commensurate magnetic correlations in in-plane Li$^{+}$ doped La$_2$CuO$_4$ are very different from the incommensurate ones in Sr$^{2+}$ or Ba$^{2+}$ doped compounds\cite{Bao2000}. The hole charges introduced by the Li$^+$ are well localized without the competing charge stripe order, however, disturb the magnetism with long range perturbations.

In this letter, we propose a constructive approach to investigate the different roles of the non-magnetic Zn$^{2+}$ and Li$^+$ impurities on the magnetic perturbations in the CuO$_2$ plane in the undoped cuprate compounds. We implement the Gutzwiller-projected spin Hall (GSH) state to model the planar Mott insulator with long range easy plane N\'eel order. The specified spin Hall (SH)  state is invariant under the time reversal and $z$-axis spin $S_z$ rotation symmetries.  The Gutzwiller projection freezes the charge degree of freedom and leads to long range spin correlations with the \textit{XY} AF spin order\cite{Ran2008,Ran2009}.   The non-magnetic impurity substitution brings zero modes for the unprojected SH state which have a local spin moment or local charge\cite{He2012}. Due to the spin $S_z$ rotation symmetry, the spin Chern number is well-defined\cite{Sheng2006,Prodan2009}. It is found that the topology described by the spin Chern number is invariant under the Gutzwiller projection. In this letter, GSH is taken as the candidate state for the spin system with easy AF order in the CuO$_2$ plane. The in-plane Zn$^{2+}$ and Li$^{+}$ substitution states are taken as the Gutzwiller-projected non-magnetic impurity doped SH states with a local spin moment for Zn$^{2+}$ and a local charged hole for Li$^{+}$, respectively. The Gutzwiller projection pushes the single particle excitation into the infinite energy region, however, preserves the local spin moment and charged hole represented by zero modes.  The spin Chern number, $C_s=2$ for all concerned states, brings the mutual Chern-Simons term that captures the relations between the charge and magnetic degrees of freedom.  While the local spin moment for Zn$^{2+}$ is screened by the long-range spin correlations, the active charge degree of freedom for Li$^{+}$ impurity generates the vortex textures which destroy the long range magnetic order with the small critical doping around $x_c\sim 0.025$.

\begin{figure}[t]
  \begin{center}
   \includegraphics[width=0.45\columnwidth]{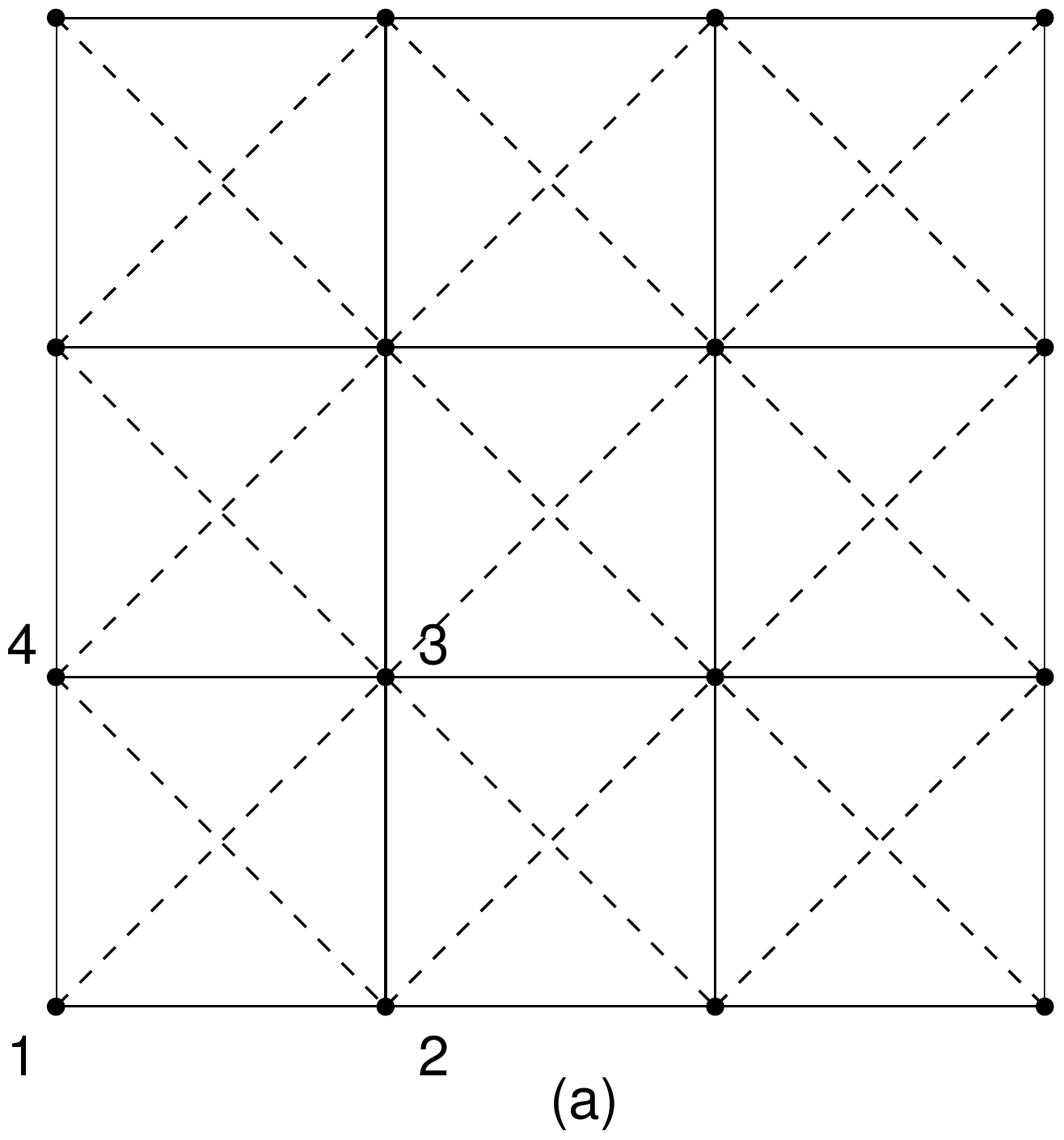}\quad\includegraphics[width=0.45\columnwidth]{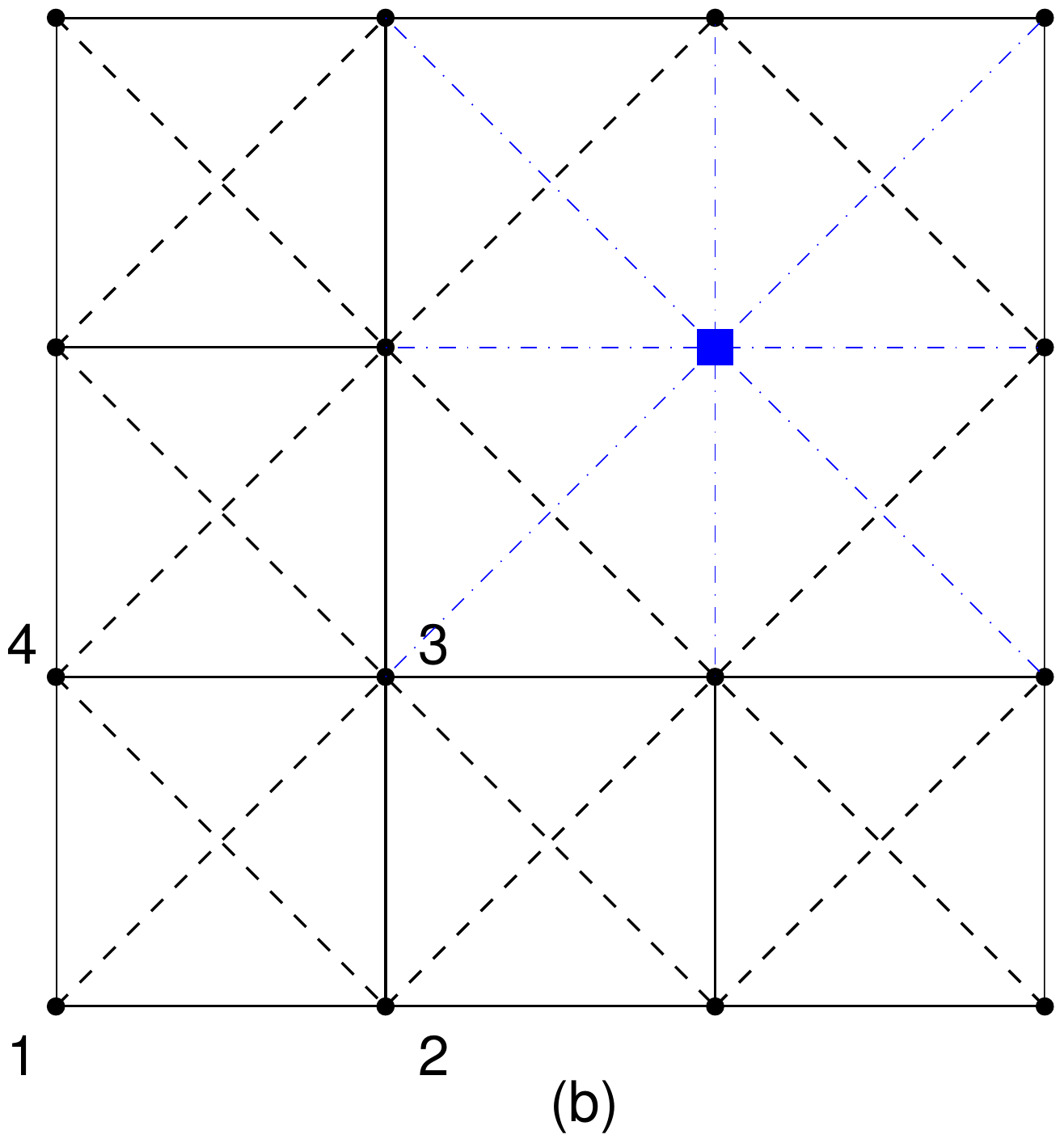}
  \end{center}
  \caption{(Color online) The square lattice without and with in-plane impurity substitution for (a) and (b), respectively.  
 The blue square marker in (b) is the vacancy prohibiting hopping process on the blue dash-dot links.}
  \label{fig:sql}
\end{figure}

We start from the unprojected SH state of the tight binding model at half filling
\begin{eqnarray}
  H_{\text{SH}}=\sum_{ij}t_{ij}^{\sigma}f_{i\sigma}^\dag f_{j\sigma},
\end{eqnarray}
with the non-zero nearest neighbor (n.n.) and next n.n. (n.n.n.) hopping parameters on the square lattice (Fig. \ref{fig:sql}). We double the unit cell by inserting uniform $\pm\pi$ flux. The hopping parameters, e.g. in the plaquette $\square_{1234}$ in Fig. \ref{fig:sql}, are chosen as the complex numbers with the fluxes as 
\begin{eqnarray}
  \phi_{\triangle_{123}}^\sigma=\phi_{\triangle_{124}}^\sigma=\phi_{\triangle_{234}}^\sigma=\phi_{\triangle_{341}}^\sigma=\sigma\frac{\pi}{2},
\end{eqnarray}
with $\phi_{\triangle_{ijk}}^\sigma=-\arg(t_{ij}^\sigma t_{jk}^\sigma t_{ki}^\sigma)$ and $\sigma=\pm1$ in formula expressions while  $\sigma=\uparrow/\downarrow$ in the index and subscript. The in-plane substitution of non-magnetic impurity prohibits the hopping process onto the spin vacancy site  which is marked as the blue square in Fig. \ref{fig:sql} (b), i.e. all the hopping amplitudes on the blue dash-dot bonds are zero.  

\begin{figure}[b]
  \begin{center}
    \includegraphics[width=0.5\columnwidth]{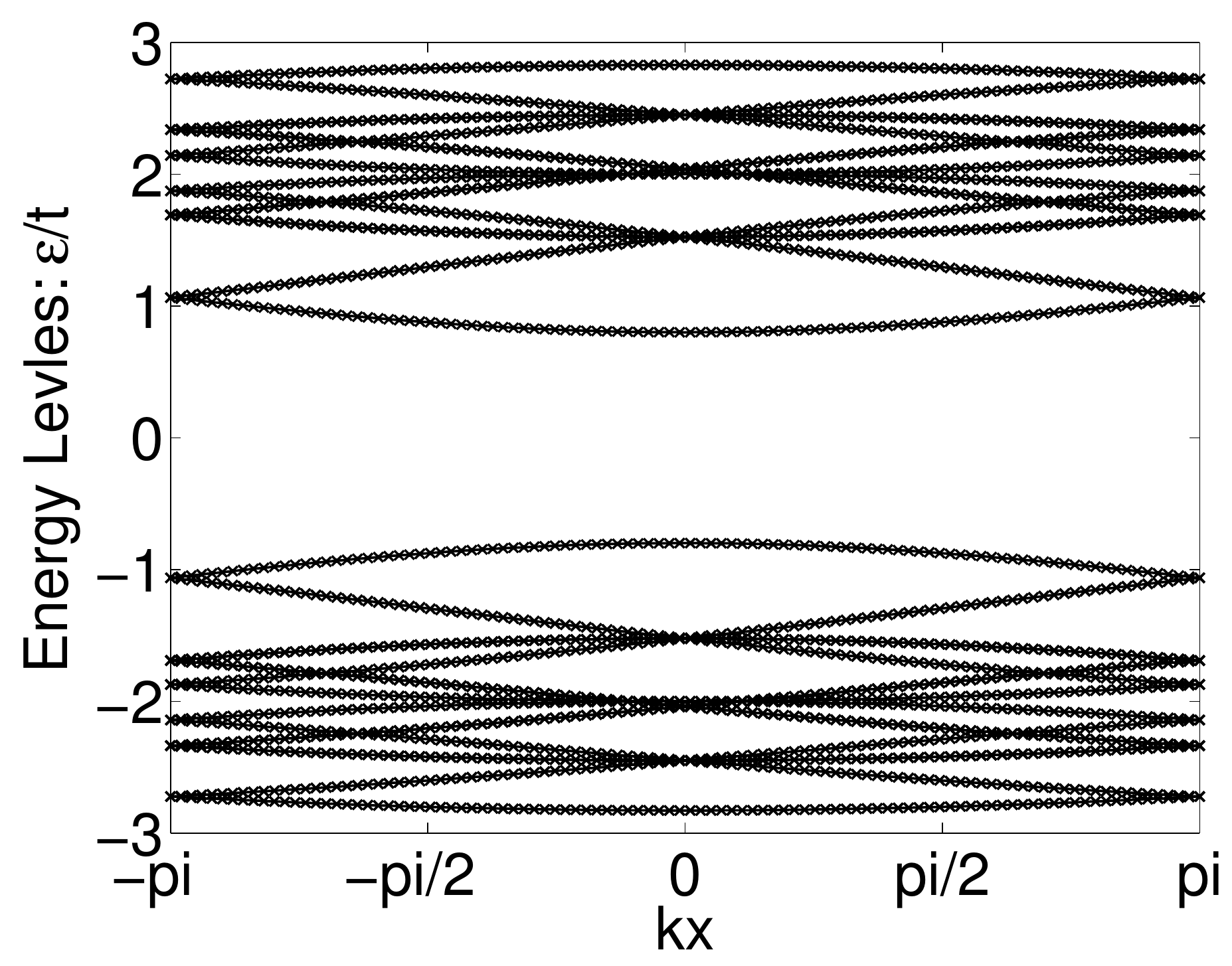}\includegraphics[width=0.5\columnwidth]{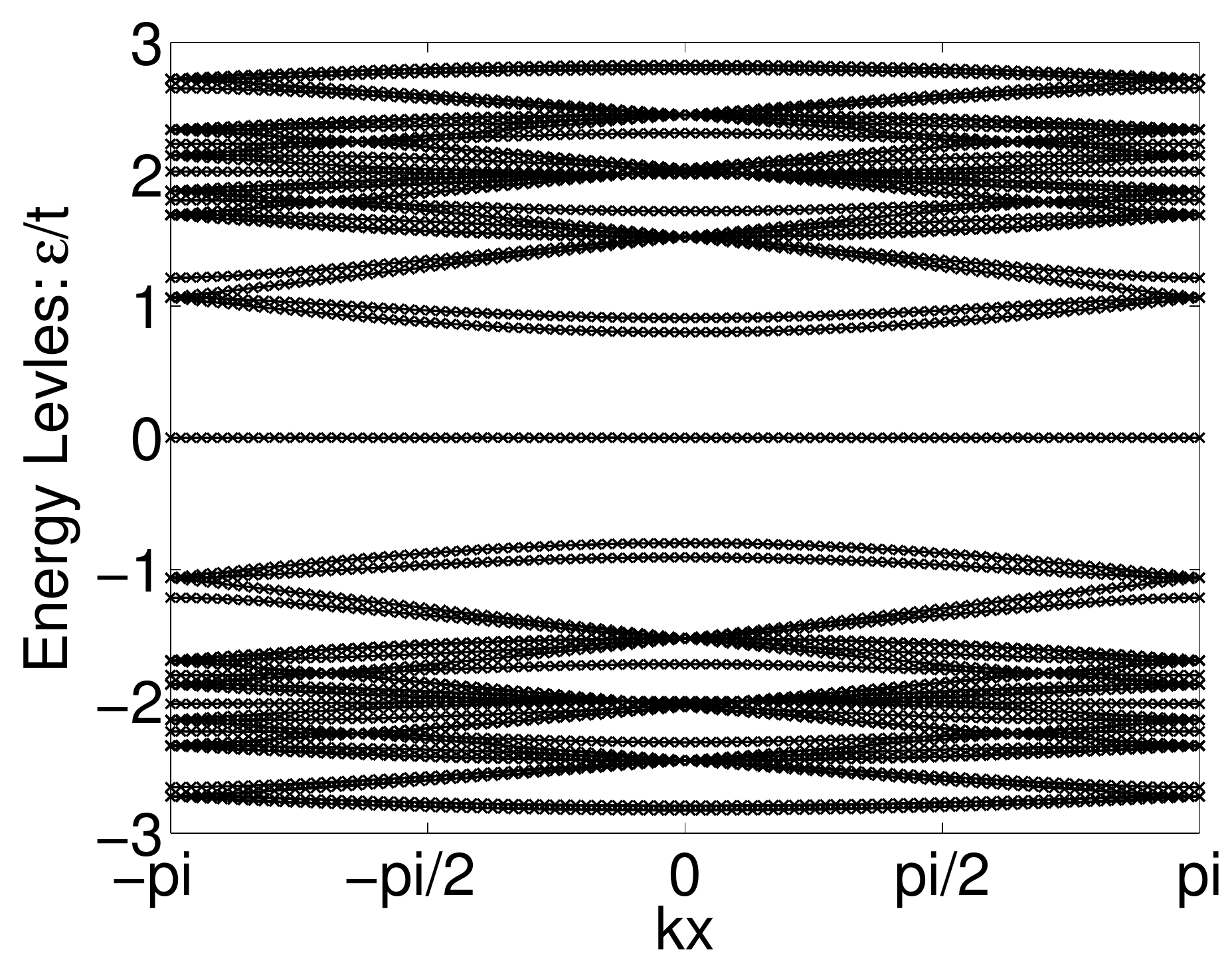}
  \end{center}
  \caption{Energy revolutions with twist boundary phase $k_y$ ($k_x=0$ fixed) for the undoped (a) and impurity doped (b) spin Hall states, respectively. }
  \label{fig:energylevel}
\end{figure}
We impose the general twisted boundary phase on the finite size system, $f_{i+L_x\sigma} = f_{i\sigma}e^{ik_x^\sigma}, f_{i+L_y\sigma}=f_{i\sigma}e^{ik_y^\sigma}$.  Under the twisted boundary conditions, the tight binding model is easily diagonalized. In this letter, the numerical calculations are carried on the square lattice on the torus with the size $L_x=L_y=8$  and the hopping amplitudes $|t|=1$ and $|t_1|=0.2|t|$ for the n.n. and n.n.n. bonds, respectively. The energy revolutions with twisted boundary phase $k_x$ ($k_y=0$ fixed) are shown in Fig. \ref{fig:energylevel} (a) and (b) for the undoped  and the in-plane impurity substitution SH model, respectively. For any boundary twist $k_x,k_y$, we can construct the undoped SH state as $|\Psi_{\text{SH}}\rangle= |\Psi_\uparrow\rangle\wedge|\Psi_\downarrow\rangle$ with $|\Psi_\sigma\rangle=|\epsilon_{\sigma}^1\rangle\wedge\cdots\wedge|\epsilon_{\sigma}^{N/2}\rangle$. Here wedge product $\wedge$ denotes the fermionic antisymmetry for the exchanges of two electrons. $\{|\epsilon_{\sigma}^i\rangle\}$ $(i=1,\cdots,N/2)$ are the negative energy levels ($N=L_x\times L_y$). The vacancy brings the zero modes that have four different vacancy doped SH states\cite{He2012}
\begin{eqnarray}
  |\Psi_{\text{SH}}^{\text{vac}};o\rangle&=& |\Psi_\uparrow^{\text{vac}}\rangle\wedge|\Psi^{\text{vac}}_\downarrow\rangle,\nonumber\\ 
  |\Psi_{\text{SH}}^{\text{vac}};\Uparrow\rangle&=& |\Psi_\uparrow^{\text{vac}}\rangle\wedge|\Psi^{\text{vac}}_\downarrow\rangle\wedge|0_\uparrow\rangle,\nonumber\\
  |\Psi_{\text{SH}}^{\text{vac}};\Downarrow\rangle&=& |\Psi_\uparrow^{\text{vac}}\rangle\wedge|\Psi^{\text{vac}}_\downarrow\rangle\wedge|0_\downarrow\rangle,\nonumber\\
  |\Psi_{\text{SH}}^{\text{vac}};e\rangle&=& |\Psi_\uparrow^{\text{vac}}\rangle\wedge|\Psi^{\text{vac}}_\downarrow\rangle\wedge|0_\uparrow\rangle\wedge|0_\downarrow\rangle,
\end{eqnarray}
with $|\Psi^{\text{vac}}_\sigma\rangle=|\epsilon_{\sigma}^1\rangle\wedge\cdots\wedge|\epsilon_{\sigma}^{N/2-1}\rangle$. The zero modes $\{|0_{\uparrow}\rangle,|0_{\downarrow}\rangle\}$ bring the charge quantum number $Q=\pm1$ ($S_z=0$) for the local charges in $|\Psi_{\text{SH}}^{\text{vac}};o/e\rangle$ and the spin quantum number $S_z=\pm1/2$ ($Q=0$) for the local spin moments in $|\Psi_{\text{SH}}^{\text{vac}};\Uparrow/\Downarrow\rangle$, respectively\cite{He2012}. 

\begin{figure}[b]
  \begin{center}
    \includegraphics[width=0.5\columnwidth]{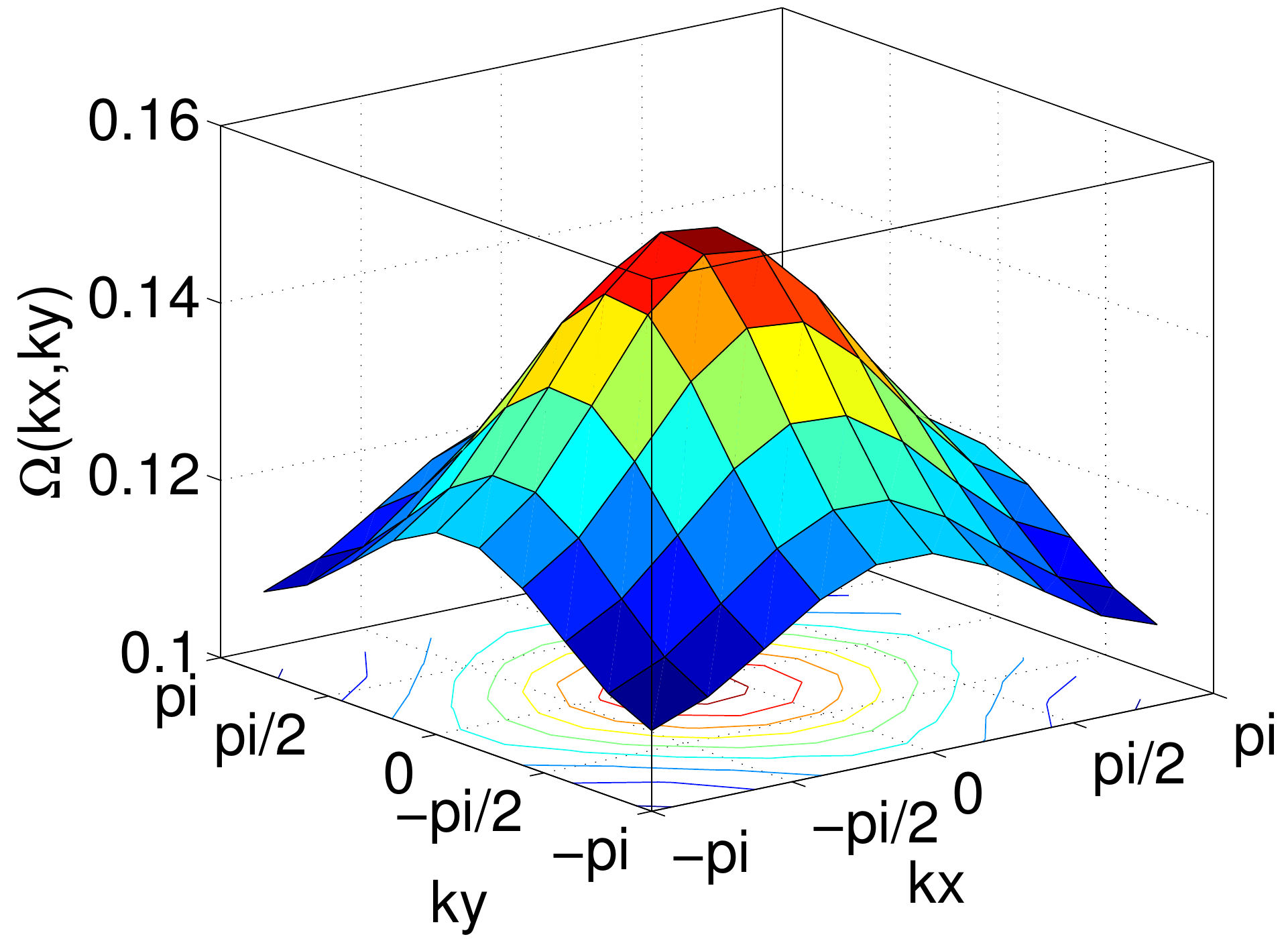}\includegraphics[width=0.5\columnwidth]{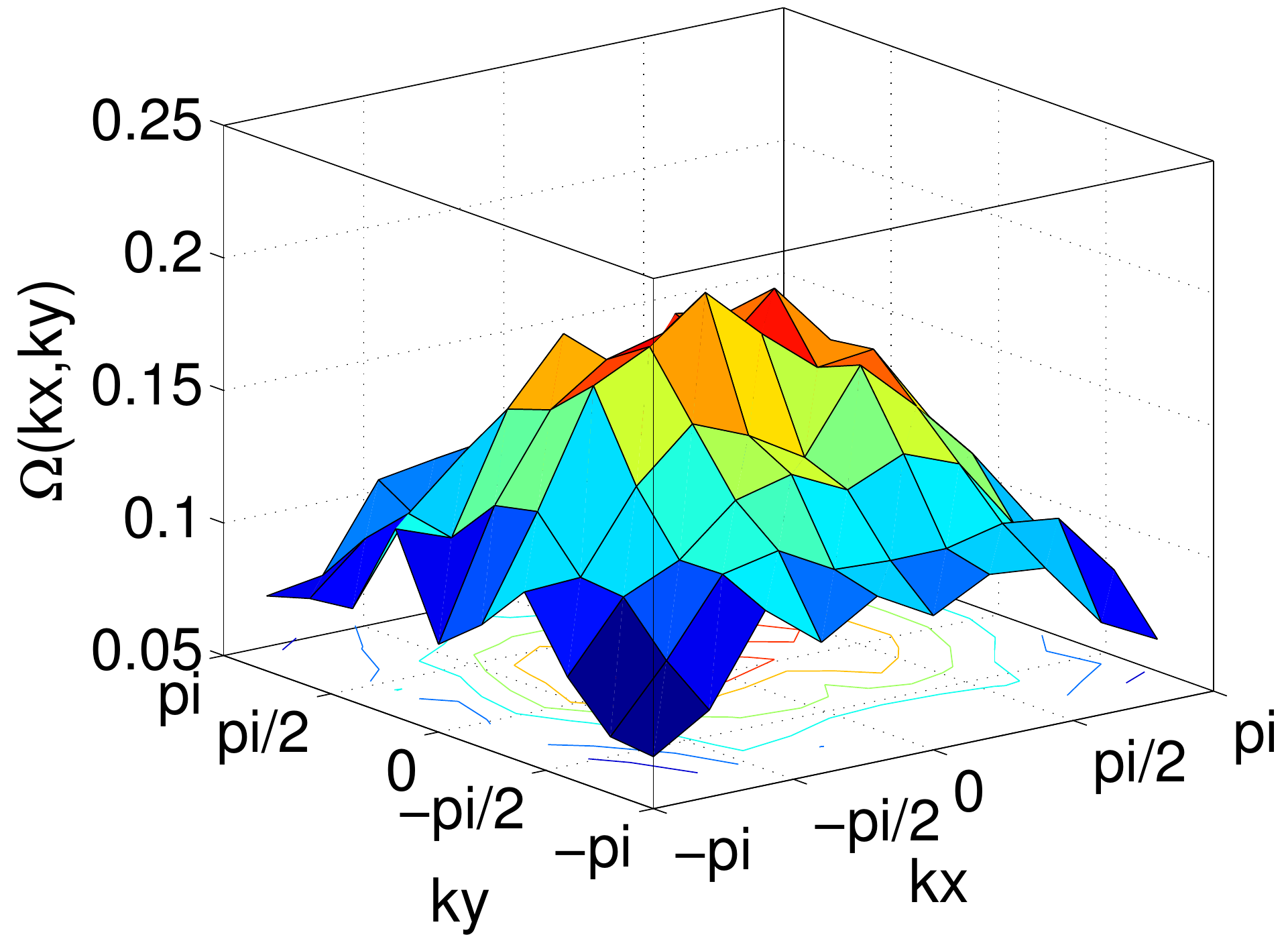}
  \end{center}
  \caption{(Color online) Spin berry curvature $\Omega(k_x,k_y)$ vs the twist $k_x$ and $k_y$ for  (a) unprojected $|\Psi_{\text{SH}}^{\text{vac}};o\rangle$, (b) projected $|\Psi_{\text{GSH}}^{\text{vac}};o\rangle$, respectively.}
  \label{fig:sbc}
\end{figure}
Since $S_z$ is conserved, it is convenient to define the spin Chern number $C_s$\cite{Sheng2006,Prodan2009}
\begin{eqnarray}
  C_s=\frac{1}{2\pi}\int_{-\pi}^{\pi} dk_x\int_{-\pi}^{\pi}dk_y \Omega(k_x,k_y),
\end{eqnarray} 
with the spin Berry curvarture $\Omega(k_x,k_y)=\nabla_{\mathbf{k}}\times\langle\Psi|i\nabla_{\mathbf{k}}|\Psi\rangle$ under the twisted boundary phase $k_x^\uparrow=-k_x^\downarrow=k_x, k_y^\uparrow=k_y^\downarrow=k_y$.  All the states have the total spin Chern number $C_{s}=2$. So the topology of the SH state is robust against the vacancy.  In Fig. \ref{fig:sbc} (a), we shown the $k$-dependent spin Berry curvatures $\Omega(k_x,k_y)$ for the state, e.g. $|\Psi_{\text{SH}};o\rangle$. It is useful to characterize the topology of the SH state in terms of the response to two different gauge fields $a^c$ and $a^s$ 
\begin{eqnarray}
  S=\int d\tau\sum_{ij\sigma}f_{i\sigma}^\dag(\partial_\tau\delta_{ij}  +t_{ij}^\sigma e^{i (\sigma\vec{a}^c+\vec{a}^s)\cdot\mathbf{r}_{ij}})f_{j\sigma},
\end{eqnarray}
which leads to the effect Lagrangian for the gauge fields
\begin{eqnarray}\label{eq:efftheo}
  \mathscr{L}_{\text{eff}}=[\frac{1}{g^2}(\epsilon_{\mu\nu\lambda}\partial_\mu(\sigma a_{\nu}^c+ a_{\nu}^s))^2+i\frac{C_{\text{s}}}{2\pi}\epsilon_{\mu\nu\lambda}a_\mu^c\partial_\nu a_{\lambda}^s].
\end{eqnarray} 
The first is the Maxwell term and the second is the mutual Chern-Simons term with the coefficient $\frac{C_{\text{s}}}{2\pi}=\frac{1}{\pi}$. As emphasized in Refs. \onlinecite{Ran2008,Ran2009}, a flux quantum of $a^s$ carries the spin quantum number $S_z=1$ and the magnetic field of $a^s$ is the $S_z$ density. Meanwhile, a flux quantum of $a^c$ will create (or annihilate) one pair of spin-up and spin-down electrons. It carries the charge quantum number $Q=2$ and the magnetic field of $a^c$ is the density of the charged pairs.

We obtain the GSH state by projecting out the double occupied state 
\begin{eqnarray}
  |\Psi_{\text{GSH}}\rangle\equiv\mathcal{P}_G|\Psi_{\text{SH}}\rangle,\quad \mathcal{P}_G\equiv\prod_i(1-n_{i\uparrow}n_{i\downarrow}).  
\end{eqnarray}
Here $\mathcal{P}_G$ is the fully Gutzwiller projection operator. The spin Chern number is unchanged under the Gutzwiller projection. The role of the projection is to freeze the charge degree of freedom of the electrons while the physics of the spin remains. Under Gutzwiller projection, $a_{\mu}^c$ is removed in $|\Psi_{\text{GSH}}\rangle$ and the spin fluctuation is controlled by the Maxwell term $a^s$. Due to the $z$-axis spin $S_z$ symmetry, the instanton effect is not important and the gauge field $a^s$ is always in the Coulomb phase and gives rise to a linear dispersed gapless photon mode corresponding to the gapless spin wave excitation. The gapless spin density fluctuations imply that $e^{i\theta S_z}$ spin rotation is spontaneously broken. So $|\Psi_{\text{GSH}}\rangle$ has the long-rang \textit{XY} spin order which has the AF pattern directly confirmed by the numerical calculation\cite{Ran2008}.

In this letter, we will take $|\Psi_{\text{GSH}}\rangle$ as the approximated candidate state for the spins in the CuO$_2$ for the cuprates. The in-plane substitution state is written as $W_{\text{vac}}^\dag\mathcal{P}_G|\Psi_{\text{SH}}\rangle$ with the vacancy inserting operator $W_{\text{vac}}^\dag$. In this paper, we assume the  permutation, i.e. $[\mathcal{P}_G,W_{\text{vac}}^\dag]=0$, so that we have the in-plane substitution state 
\begin{eqnarray}
  |\Psi_{\text{GSH}}^{\text{vac}};*\rangle\equiv\mathcal{P}_G|\Psi_{\text{SH}}^{\text{vac}};*\rangle,
\end{eqnarray}
with $*\in\{o,\Uparrow,\Downarrow\}$. The Gutzwiller projection pushes the single particle excitation into infinite energy region, however, preserves the local spin moment and hole in  $|\Psi_{\text{SH}}^{\text{vac}};\Uparrow,\Downarrow,o\rangle$. So the permutation assumption, i.e. $[\mathcal{P}_G,W_{\text{vac}}^\dag]=0$, is quite reasonable. We will study the in-plane substitution states $|\Psi_{\text{GSH}}^{\text{vac}};\Uparrow\rangle$ and $|\Psi_{\text{GSH}}^{\text{vac}};\Downarrow\rangle$ to investigation the magnetic disturbances for Zn$^{2+}$ and $|\Psi_{\text{GSH}}^{\text{vac}};o\rangle$ for Li$^+$, respectively. The projected states have the same spin Chern number as unprojected ones, i.e. $C_s=2$. In Fig. \ref{fig:sbc} (b), we show the $k$-dependent spin Berry curvature for the projected state, e.g. $|\Psi_{\text{GSH}}^{\text{vac}};o\rangle$. 

\begin{figure}[b]
  \begin{center}
    \includegraphics[width=0.5\columnwidth]{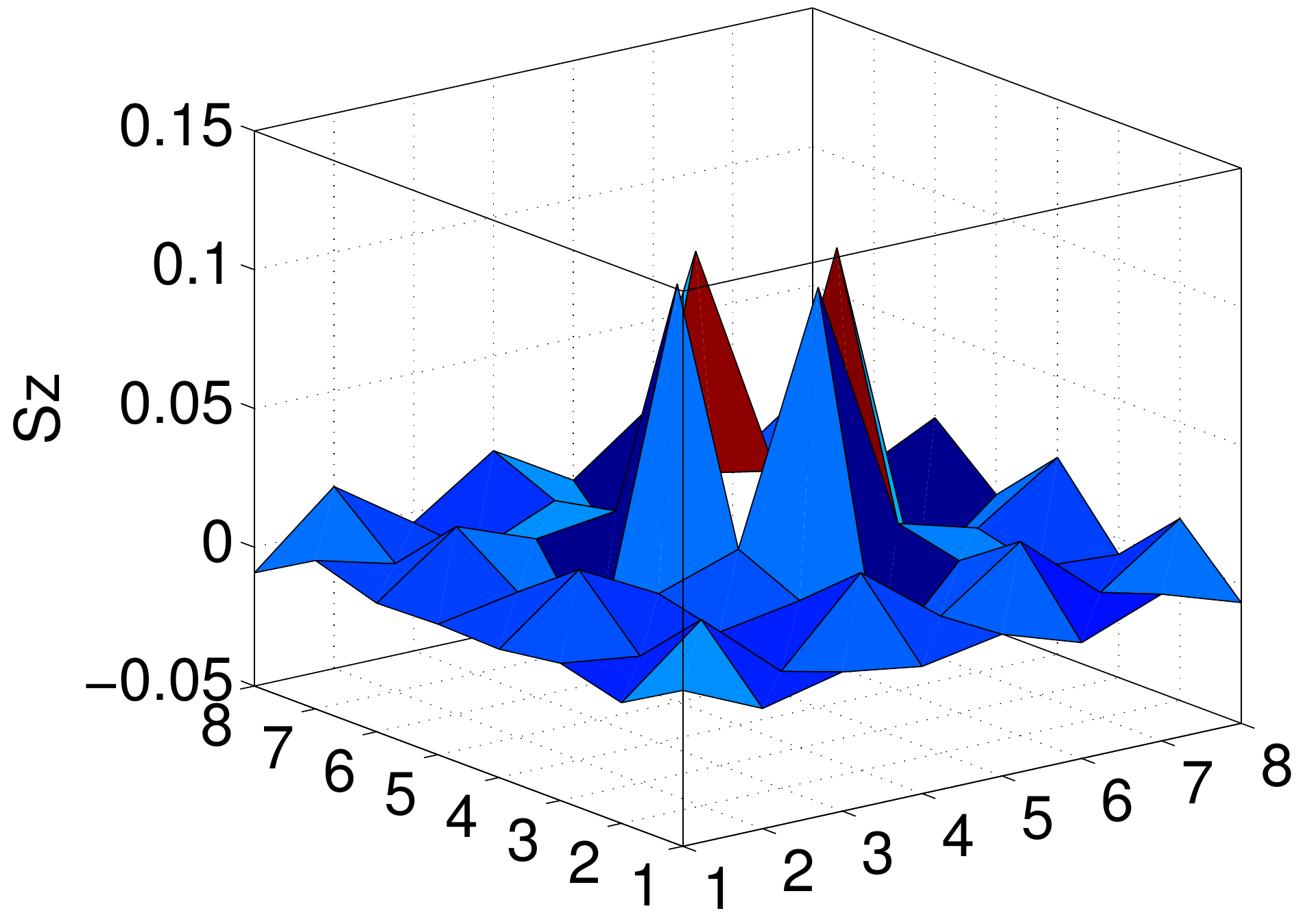} 
  \end{center}
  \caption{The spin $S_z$ distribution for the in-plane substitution state $|\Psi_{\text{GSH}}^\text{vac};\Uparrow\rangle$.}
  \label{fig:spinzn}
\end{figure}
The Zn$^{2+}$ in-plane substitution states, i.e. $|\Psi_{\text{GSH}}^\text{vac};\Uparrow\rangle$ and $|\Psi_{\text{GSH}}^{\text{vac}};\Downarrow\rangle$, have no charge degree of freedom after the Gutzwiller projection, i.e. $Q=0$. The zero modes brings a local moment with the spin $S_z$ distribution shown in Fig. \ref{fig:spinzn} for $|\Psi_{\text{GSH}}^\text{vac};\Uparrow\rangle$. It is localized around the spin vacancy in the staggered pattern arising from the Gutzwiller projection. In the thermodynamical limit, the spin fluctuation is described by the dynamics of the gauge fields $a^s$ in the Maxwell term
\begin{eqnarray}
  \mathscr{L}_{\text{MW}}=\frac{1}{g^2}(\epsilon_{\mu\nu\lambda}\partial_{\mu}a_{\nu}^s)^2,
\end{eqnarray}
with the gapless excitation implying the long range \textit{XY} spin order. In the continuum limit, we have the static equations of motion
\begin{eqnarray}
  \nabla^2\vec{a}^s=0,\quad\nabla\times\vec{a}^s(\mathbf{r})=\hat{z}2\pi(S_{\text{loc}}^z(\mathbf{r})+S_{\text{in}}^z(\mathbf{r})),
\end{eqnarray}
with $a_0^s=0$. $S_{\text{loc}}^z(\mathbf{r})$ is the spin $S_z$ distribution as shown in Fig. \ref{fig:spinzn} and can be approximated as $S_{\text{loc}}^z(\mathbf{r})=\frac{1}{2}\delta(\mathbf{r})$ in the continuum limit. $S_{\text{in}}^z(\mathbf{r})$ is the induced spin configuration that screens the local moment. When S$^z$ excitations are gapless, we can obtain the full screening for the local moment resembling the Kondo screening for the local moment in the metal.  The local moment will survive provided that the spins are short range correlated and the screening is not complete. This may be the reason for the local moment associated with the dopant Zn$^{2+}$ in the pseudogap phase with short range spin order in the underdoped cuprate compounds. Regardless of the screening, the dopant Zn$^{2+}$ brings only the short range magnetic disturbance and doesn't change the \textit{XY} AF spin order in $|\Psi_{\text{GSH}}^{\text{vac}};\Uparrow/\Downarrow\rangle$. It behaves as the site dilution percolation.

\begin{figure}[b]
  \begin{center}
    \includegraphics[width=0.5\columnwidth]{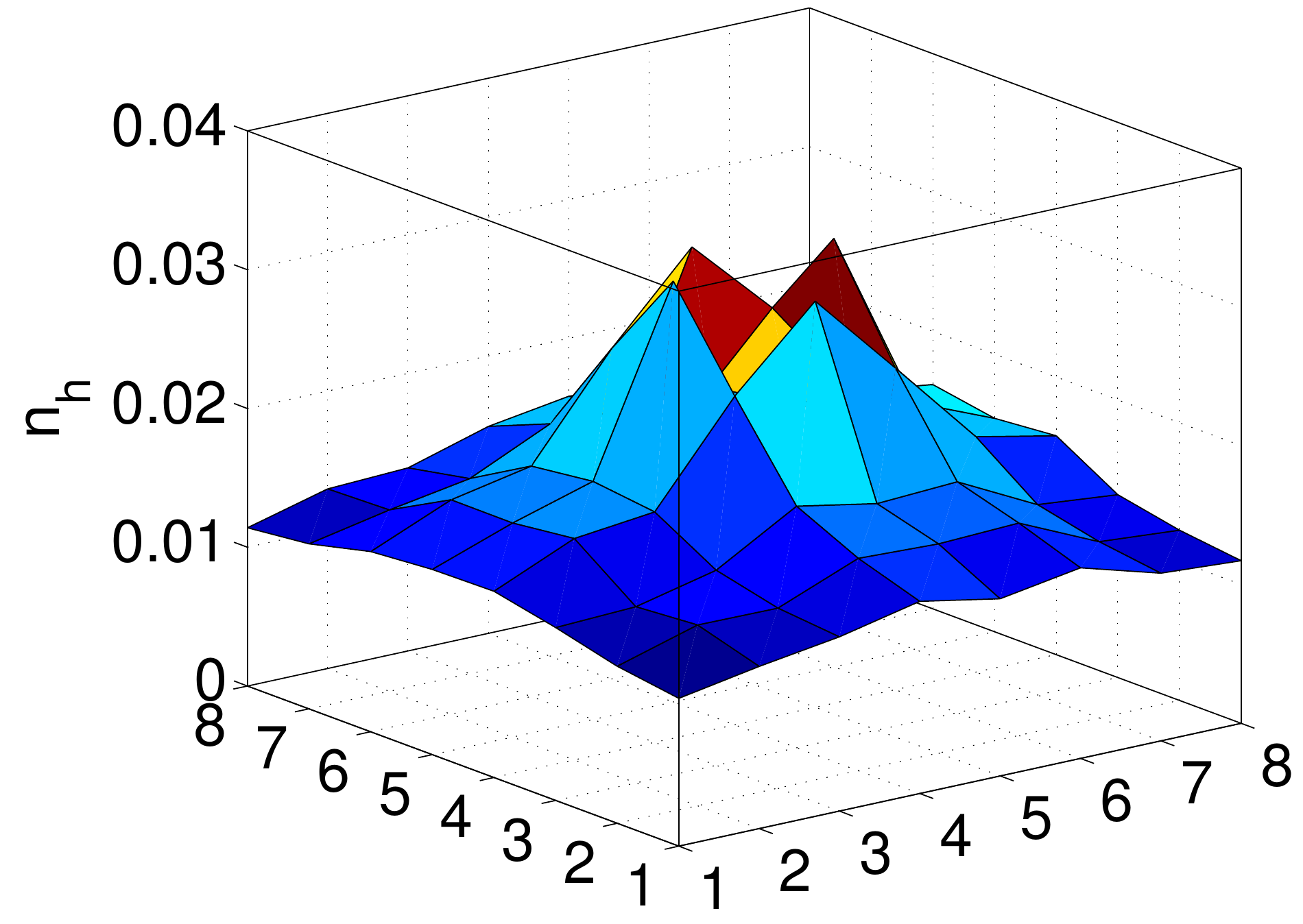}\includegraphics[width=0.5\columnwidth]{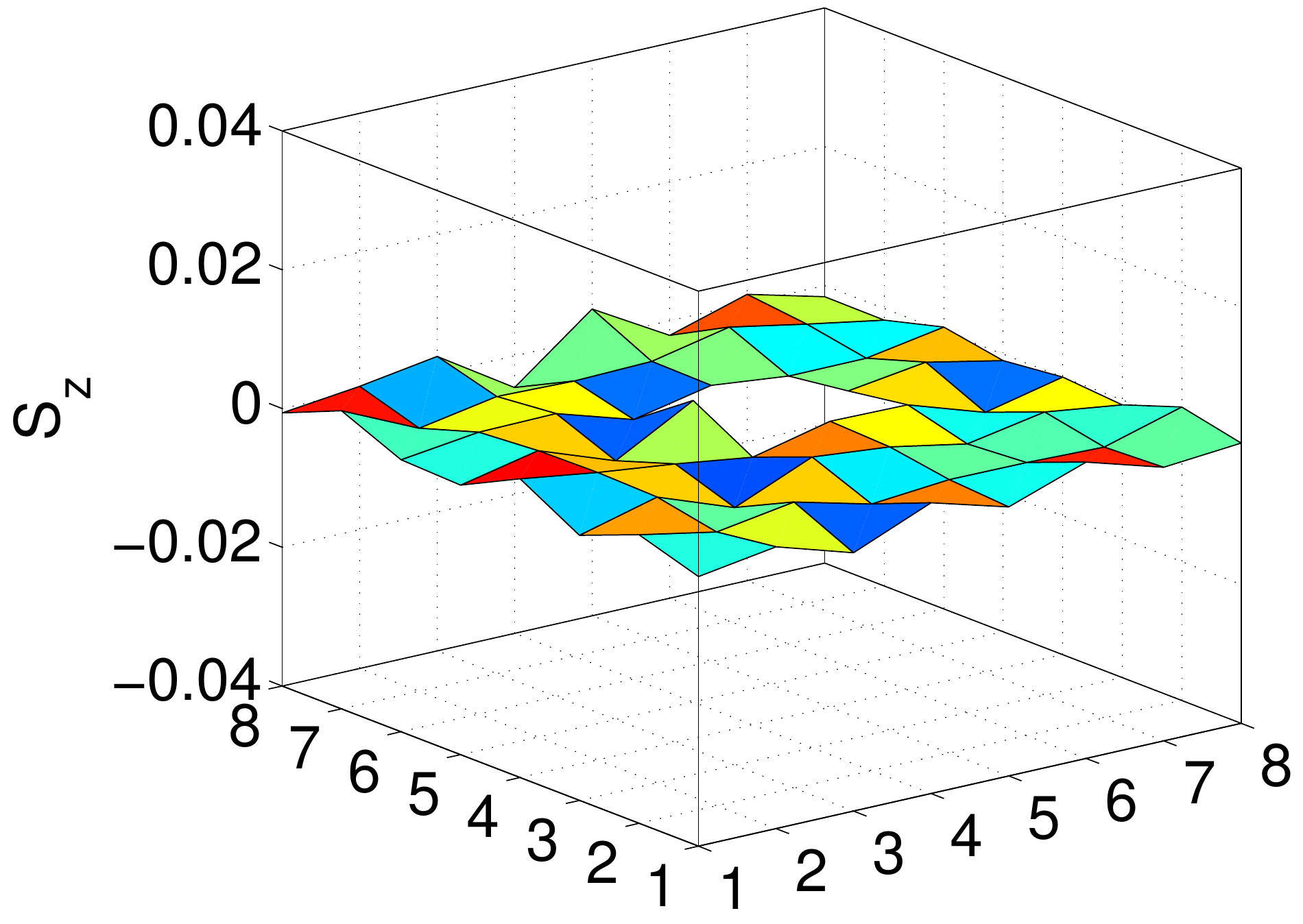}
  \end{center}
  \caption{(a) The hole $n_h$ and (b) spin $S^z$ distribution for $|\Psi_{\text{GSH}}^{\text{vac}};o\rangle$.}
  \label{fig:disli}
\end{figure}
Comparing with Zn$^{2+}$, the hole in the in-plane substitution state of Li$^+$, i.e. $|\Psi_{\text{GSH}}^{\text{vac}};o\rangle$, replaces the local moment in $|\Psi_{\text{GSH}}^{\text{vac}};\Uparrow/\Downarrow\rangle$. The distributions of the hole and spin for $|\Psi_{\text{GSH}}^{\text{vac}};o\rangle$ are shown in  Fig. \ref{fig:disli} (a) and (b), respectively. Due to the correlated effect of the Gutzwiller projection, the distributions vary from $S^z(\mathbf{r})$ for $|\Psi_{\text{GSH}}^{\text{vac}};\Uparrow\rangle$ shown in Fig. \ref{fig:spinzn} to $n_h(\mathbf{r})$ for $|\Psi_{\text{GSH}}^{\text{vac}};o\rangle$ in Fig. \ref{fig:disli} (a).  It is intriguing that the charge degree of freedom  for Li$^+$ activates the mutual Chern-Simons term 
\begin{eqnarray}\label{eq:mcs}
  \mathscr{L}_{\text{MCS}}= \frac{i}{\pi}\epsilon_{\mu\nu\lambda}a_{\mu}^c\partial_\nu a_{\lambda}^s.
\end{eqnarray}
In the continuum limit, we have the equations of motion 
\begin{eqnarray}
  \nabla^2a_0^s=n_h(\mathbf{r}),\quad \vec{a}^s=0,
\end{eqnarray}
where $n_h(\mathbf{r})=\frac{1}{\pi}\epsilon_{0\mu\nu}\partial_{\mu}a_{\nu}^c$ is the charge distribution of the hole in $|\Psi_{\text{GSH}}^{\text{vac}};o\rangle$ in Fig. \ref{fig:disli} (a).  There is no induced charge that screens $n_h(\mathbf{r})$.  Due to the mutual Chern-Simons term, the charged hole generates the $\pi$ and $-\pi$ fluxes in $f_{i\uparrow}$ and $f_{i\downarrow}$, respectively. Therefore, Li$^{+}$ generates a $2\pi$ vortex  in the XY spin order $S^+=f_{i\uparrow}^\dag f_{i\downarrow}$. Considering the compactness, the gauge field $a^c$ should be written as
\begin{eqnarray}
  a_\mu^c\rightarrow a_\mu^c+2\pi\mathcal{N}_{\mu},\quad n_h(\mathbf{r})\rightarrow n_h(\mathbf{r})+2n,
\end{eqnarray}
with the integer field $\mathcal{N}_{\mu}$ ($n\in\mathbb{Z}$) that keeps track of the compactness.  When $n=-1$, the integer field $\mathcal{N}_\mu$ can convert the $2\pi$ vortex in \textit{XY} spin order into $-2\pi$ anti-vortex.

\begin{figure}[b]
  \begin{center}
    \includegraphics[width=0.45\columnwidth]{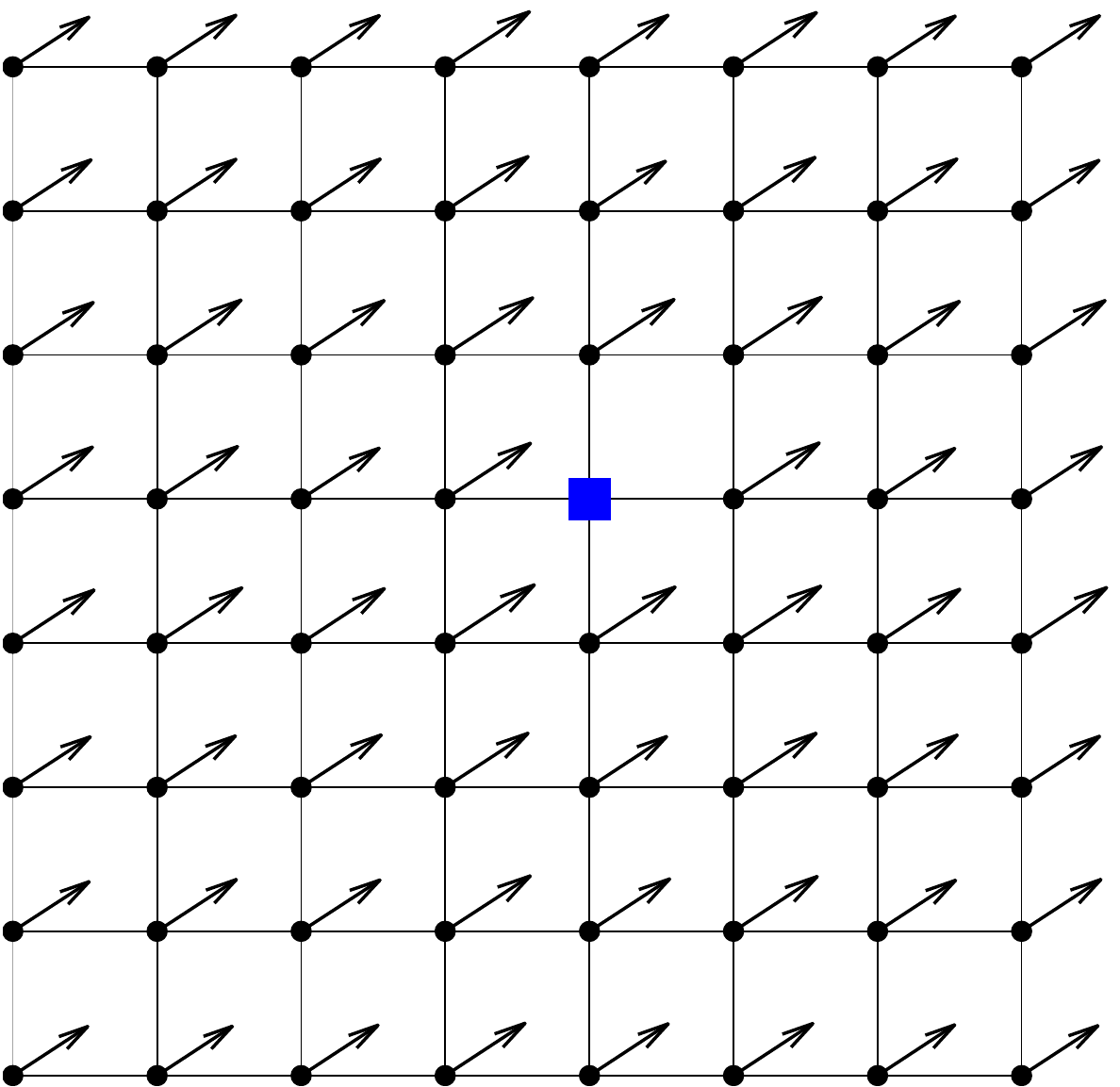}\quad\includegraphics[width=0.42\columnwidth]{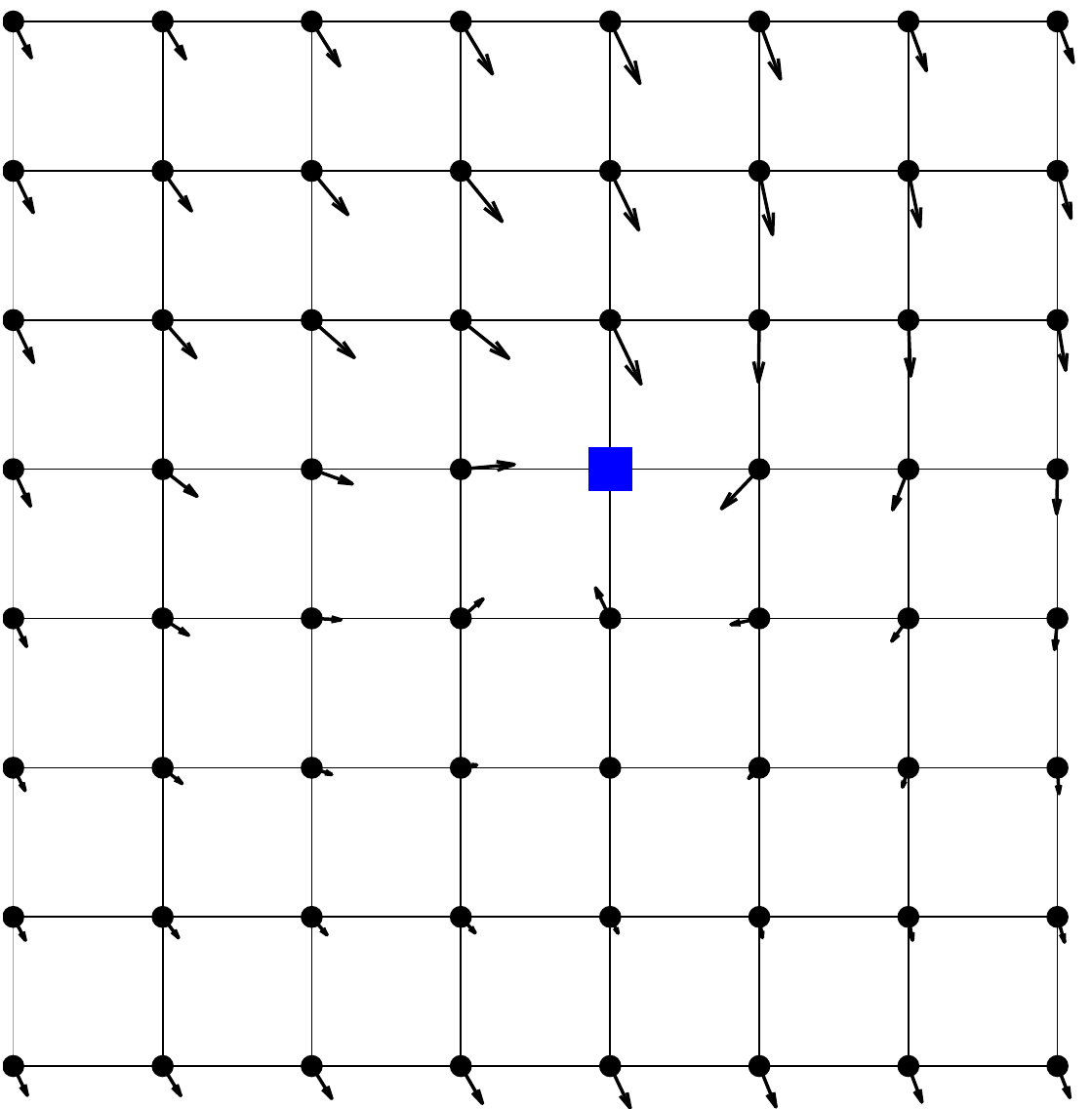}
  \end{center}
  \caption{ XY AF spin configuration $M_{\text{AF}}(\mathbf{r}_i)$ for (a). Zn$^{2+}$ substitution state $|\Psi_{\text{GSH}}^{\text{vac}};\Uparrow\rangle$, (b). Li$^{+}$ substitution state $|\Psi_{\text{GSH}}^{\text{vac}};o\rangle$, respectively. The AF prefactor $(-1)^{\mathbf{r}_i}$ is already included in the definition of $M_{\text{AF}}(\mathbf{r}_i)$, e.g. in Eq. (\ref{eq:af}). In (b), the imposed periodic boundary condition on the torus generates the anti-vortex paired with the vortex generated by the Li$^+$ impurity}
  \label{fig:xy}
\end{figure}

From the mutual Chern-Simons term (\ref{eq:mcs}), we know that Zn$^{2+}$ does little damage on the XY AF spin order while Li$^+$ generates the destructive vortex structure. Below we shall confirm the results by numerical computation directly. For the finite system, we have the expression for the AF spin order\cite{Ran2008,Ran2009}
\begin{eqnarray}\label{eq:af}
 M_{\text{AF}}=(-1)^{\mathbf{r}_i}\langle\Psi_{\text{GSH}}|S^+_i|\Psi_{\text{GSH,(-1)-flux}}\rangle,
\end{eqnarray}
with the projected \textit{minus} 1-flux state  $|\Psi_{\text{GSH,(-1)-flux}}\rangle$. $(-1)^{\mathbf{r}_i}$ is the AF prefactor. Without in-plane substitution, the spin configuration is in the AF N\'eel patter, $M_{\text{AF}}\sim \text{const.}$\cite{Ran2008}.  For Zn$^{2+}$ substitution, the AF spin order for $|\Psi_{\text{GSH}}^{\text{vac}};\Uparrow\rangle$ is given as
\begin{eqnarray}
  M_{\text{AF}}^{\text{Zn}}=(-1)^{\mathbf{r}_i}\langle \Psi_{\text{GSH}}^{\text{vac}};\Uparrow|S_i^+|\Psi_{\text{GSH,(-1)-flux}}^{\text{Zn}}\rangle,
\end{eqnarray}
with the projected minus 1-flux state
\begin{eqnarray}
  |\Psi_{\text{GSH,(-1)-flux}}^{\text{Zn}}\rangle=\mathcal{P}_G|\Psi_{\uparrow;\text{Zn}}^{\text{(-1)-flux}}\rangle\wedge|\Psi_{\downarrow;\text{Zn}}^{\text{(-1)-flux}}\rangle,
\end{eqnarray}
with $|\Psi_{\uparrow;\text{Zn}}^{\text{(-1)-flux}}\rangle=|e_{\uparrow}^1\rangle\wedge\cdots\wedge|e_{\uparrow}^{N/2-1}\rangle$ and $|\Psi_{\downarrow;\text{Zn}}^{\text{(-1)-flux}}\rangle=|e_{\downarrow}^1\rangle\wedge\cdots\wedge|e_{\downarrow}^{N/2}\rangle$. $|e_{\sigma}^i\rangle$ is the negative energy level in the presence of the uniform flux integrated to a minus flux quantum.  Similarly, the XY AF order for Li$^{+}$ substitution state $|\Psi_{\text{GSH}}^{\text{vac}};o\rangle$ is given as
\begin{eqnarray}
  M_{\text{AF}}^{\text{Li}}=(-1)^{\mathbf{r}_i}\langle \Psi_{\text{GSH}}^{\text{vac}};o|S_i^+|\Psi_{\text{GSH,(-1)-flux}}^{\text{Li}}\rangle,
\end{eqnarray}
with the projected minus 1-flux impurity doped SH state
\begin{eqnarray}
  |\Psi_{\text{GSH,(-1)-flux}}^{\text{Li}}\rangle=\mathcal{P}_G|\Psi_{\uparrow;\text{Li}}^{\text{(-1)-flux}}\rangle\wedge|\Psi_{\downarrow;\text{Li}}^{\text{(-1)-flux}}\rangle
\end{eqnarray}
with $|\Psi_{\uparrow;\text{Li}}^{\text{(-1)-flux}}\rangle=|e_{\uparrow}^1\rangle\wedge\cdots\wedge|e_{\uparrow}^{N/2-2}\rangle$ and $|\Psi_{\downarrow;\text{Li}}^{\text{(-1)-flux}}\rangle=|e_{\downarrow}^1\rangle\wedge\cdots\wedge|e_{\downarrow}^{N/2}\rangle$. 

In Fig. \ref{fig:xy}, we show the numerical results of XY AF spin order for Zn$^{2+}$ and Li$^{+}$ substitution states in (a) and (b), respectively. As expected, $|\Psi_{\text{GSH}}^{\text{vac}};\Uparrow\rangle$ has the untwisted XY AF spin order while $|\Psi_{\text{GSH}}^{\text{vac}};o\rangle$ has the twisted spin order with the vortex structure. It should be noted that the imposed periodic boundary condition on the torus generates the anti-vortex paired with the vortex generated by the Li$^+$ impurity. 

The vortex structure generated by Li$^{+}$ substitution can destroy the long range spin order due to the Kosterlitz-Thouless transition\cite{Kosterlitz1973}. In the presence of the vortex-antivortex pairs, the renormalized spin stiffness at temperature $T$ is given as\cite{Chaikin2000,Timm2000,Mei2010}
\begin{eqnarray}
  \frac{\rho_s^R}{T}=\frac{\rho_s}{T}-4\pi^3 y^2 (\frac{\rho_s}{T})^2\int_{r_0}^\infty \frac{dr}{r_0}\left( \frac{r}{r_0} \right)^{3-2\pi\rho_s/T},
\end{eqnarray}
with $\rho_s$ is the bare stiffness without vortex and $y^2=y^2_{\text{Li}}+y^2_{\text{th}}$ are the total fugacity of vortex pairs where $y^2_{\text{Li}}$ for the pairs generated by Li$^{+}$ and $y^2_{\text{th}}$ for the thermally generated pairs.  $r_0$ is the size of the vortex core. At the critical temperature $T_c$, we have $\rho_s^R=2T_c/\pi$.  The rigidity of the spin order can only sustain the amount of the vortex-antivortex pairs with the fugacity\cite{Timm2000,Mei2010}
\begin{eqnarray}
  y^2_{\text{Li}}+y^2_{\text{th}}=\frac{1}{2\pi^2}(1-\frac{2T_c}{\rho_s\pi})^2.
\end{eqnarray}
At the critical concentration of Li$^+$, $T_c=0$ and $y^2_{\text{th}}=0$ such that $y^2_{\text{Li}}=\frac{1}{2\pi^2}$ and we obtain the critical concentration
\begin{eqnarray}
  x_c=2y^2_{\text{Li}}({a}/{r_0})^2\simeq\frac{1}{4\pi^2}=0.025,
\end{eqnarray}
with the estimation $r_0=2a$. 

During our construction, all the unprojected states are protected by the finite gaps. The spin Chern numbers of these states are well-defined and remain unchanged after projection.  The key point in our construction is the nonzero spin Chern number which leads to the mutual Chern-Simons term when the charge degree of freedom is activated. The mutual Chern-Simons structure in the doped antiferromagnet is extensively studied for the cuprate by Weng's group starting from the bosonic resonant valence state\cite{Kou2005,Mei2010,Ye2011,Ye2012}. The spin and charge texture for the substitution  of Li$^{+}$ in the CuO$_2$ plane was also studied in Ref. \onlinecite{Haas1996} where the skyrmion structure was proposed. The spin texture induced by the delocalized hole was also studied in the the eary literatures of high $T_c$ cuprates\cite{Shraiman1988,Shraiman1990}.

In conclusion, we present a constructive approach to investigate the magnetic disturbance for the in-plane substitutions for Zn$^{2+}$ and Li$^+$ for Cu$^{2+}$ in the CuO$_2$ planes in the undoped cuprate compound. In the substitution states,  Zn$^{2+}$ and Li$^{+}$ impurities can be represented as vacancies introducing a zero mode, which has a local spin moment for Zn$^{2+}$ and a charged hole for Li$^{+}$, respectively. While the local spin moment for Zn$^{2+}$ is screened by the long-range AF spin correlations, the active charge degree of freedom for Li$^{+}$ impurity generates the vortex in the spin AF background and suffices to suppress the spin order at the critical concentration $x_c\sim0.025$.

During the work, from the issue raising to the manuscript preparing, many helpful suggestions and comments have been given by T.M. Rice. JW Mei also thanks M. Sigrist and Z.Y. Weng for useful discussions.  The suggestion of numerical methods from Lei Wang is also useful. The work is supported by Swiss NationalFonds. 

\bibliography{vacancy}
\end{document}